\documentclass{aa}

\usepackage{graphicx}
\usepackage{txfonts}

\begin{document}

\title{H$\alpha$ variability of the recurrent nova T Coronae Borealis\thanks{based
on observations obtained at Rozhen National Astronomical Observatory, Bulgaria}}

   \author{V. Stanishev \inst{1,}\thanks{now at Physics Department, Stockholm University}
              \and
           R. Zamanov \inst{2}
              \and 
	   N. Tomov\inst{3}
               \and
	   P. Marziani \inst{4}
      }

   \institute{Institute of Astronomy, Bulgarian Academy of Sciences, 
    72 Tsarigradsko Shouse Blvd., 1784 Sofia, Bulgaria
   \and
     Astrophysics Research Institute, Liverpool John Moores University, 
     Twelve Quays House, Egerton Wharf, Birkenhead CH41 1LD, UK
  \and
    Institute of Astronomy and Isaac Newton Institute of Chile - Bulgarian
    Branch, National Astronomical Observatory Rozhen, PO Box 136, 4700 Smolyan,
    Bulgaria
  \and
    INAF, Osservatorio Astronomico di Padova, Vicolo dell'Osservatorio 5,
    35122 Padova, Italy
}

   \offprints{V. Stanishev (\email{vall@astro.bas.bg})}

   \date{Received ;accepted}

   \abstract{
We analyze H$\alpha$ observations of the recurrent nova  \object{T
CrB} obtained during the last decade. For the first time the
H$\alpha$ emission profile is analyzed after subtraction of
the red giant contribution. Based on our new radial velocity
measurements of the H$\alpha$ emission line we estimate the component
masses of \object{T~CrB}. It is found that the hot component 
is most likely a massive white dwarf. We estimate the inclination 
and the component masses to be
$i\simeq67^\circ$, $M_{\rm WD}\simeq1.37\,\pm0.13M_\odot$ and
$M_{\rm RG}\simeq1.12\,\pm0.23M_\odot$, respectively. The radial
velocity of the central dip in the H$\alpha$ profile changes nearly in
phase with that of the red giant's absorption lines. This suggests
that the dip is most likely produced by absorption 
in the giant's wind. 

Our observations cover an interval when the
H$\alpha$ and the $U$-band flux vary by a factor of $\sim6$, while
the variability in $B$ and $V$ is much smaller.
Based on our observations, and archival
ultraviolet and optical data we show that the optical, ultraviolet 
and H$\alpha$ fluxes strongly correlate.
We argue that the presence of an accretion disc can account for most of 
the observed properties of \object{T~CrB}.

\keywords{accretion, accretion discs -- stars: individual:
\object{T~CrB} -- novae, cataclysmic variables -- binaries: symbiotic}
   }
 \authorrunning{V. Stanishev et al.}
 \titlerunning{H$\alpha$ variability of the recurrent nova T Coronae Borealis}
 \maketitle

\section{Introduction}

\object{T Coronae Borealis} is a well known recurrent nova with an orbital
period of 227\fd57 (Kenyon \& Garcia \cite{kg}; Fekel et al.
\cite{fek}). The system contains  a M4.5\,III giant 
(Murset \& Schmid \cite{mur}), which fills
its Roche lobe and transfers mass toward a hot component (HC).
The optical spectrum of
\object{T~CrB} is dominated by the red giant (RG) with superimposed Balmer
emission lines, which are thought to originate in an accretion
disc (AD) around the HC (Selvelli et al. 1992; hereafter
\cite{sel}). The high excitation \ion{He}{ii}\,$\lambda$4686 emission
line is sometimes also visible (Iijima \cite{iij}). The H$\alpha$
emission line is well visible all the time and shows variability
with the orbital period as well as large variations on a time scale
of years (Anupama \& Prabhu \cite{anu}; Zamanov \& Marti
\cite{zm}). The photometric variability in the long wavelength
range is dominated by the ellipsoidal variability of the RG (Shahbaz et al.
\cite{sha}; Belczynski \& Mikolajewska \cite{bm}; Yudin \& Munari \cite{yum}). 
At short wavelengths \object{T~CrB} shows strong flickering activity,
with the same parameters as those of normal
cataclysmic variables, in spite of the large difference in the size
of the AD (Zamanov \& Bruch \cite{zam}). Apart from
the recurrent nova outbursts, \object{T~CrB} shows other periods
of enhanced luminosity. The amplitude of these variations
increases toward the shorter wavelengths and reaches $\sim2$ mag in
the $U$-band (Zamanov \& Zamanova \cite{zz}; Hric et al. \cite{hric}).
The equivalent width of the line H$\alpha$
(EW(H$\alpha$)) also greatly increases  and reaches 
$\sim30-35$ \AA\  (Anupama \& Prabhu \cite{anu}; 
Zamanov \& Marti \cite{zm}). This activity is not well understood.

 \begin{figure*}[t]
 \includegraphics*[width=17cm]{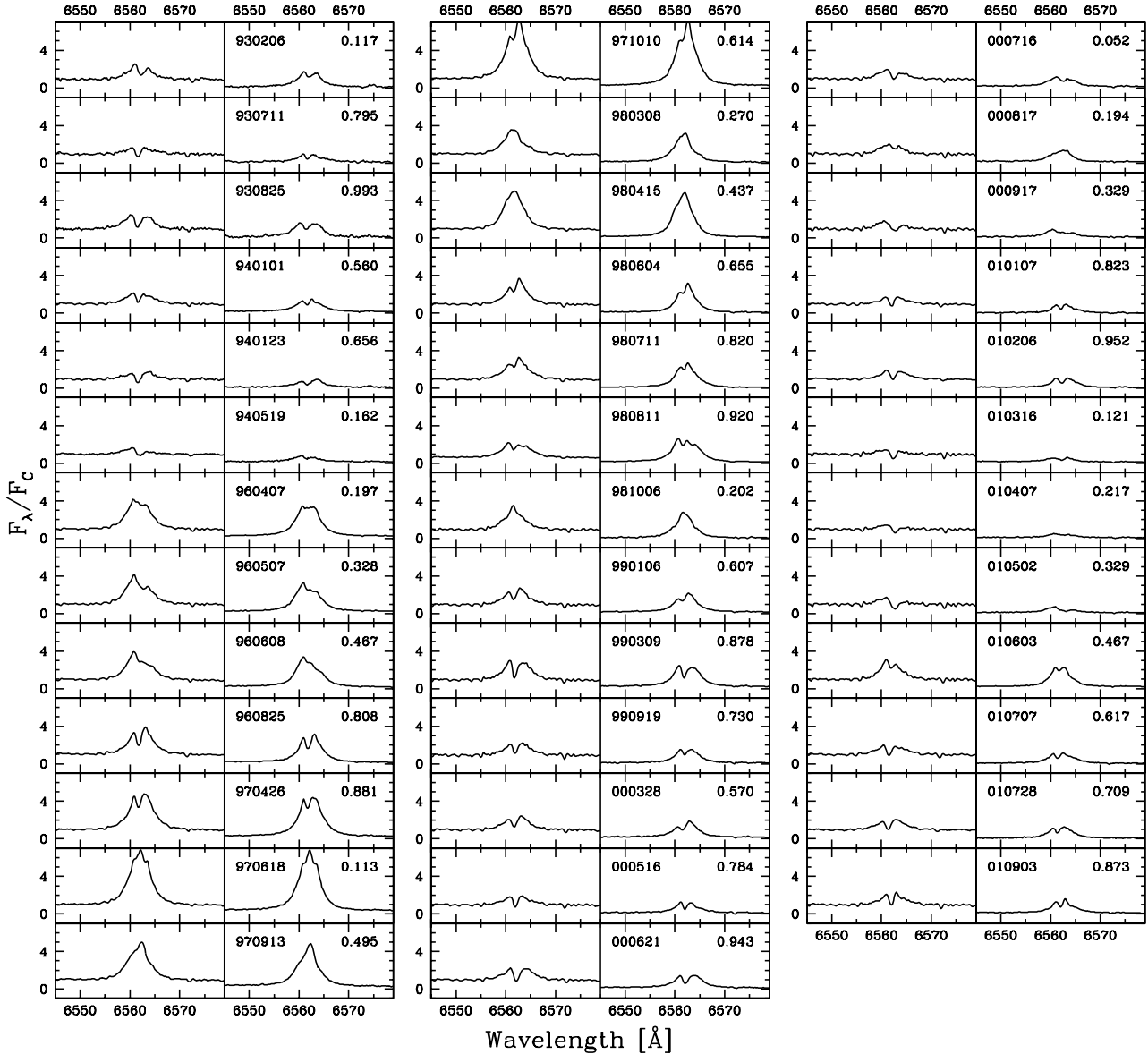}
 \caption[]{H$\alpha$ emission line profiles before (left panels)
 and after (right panels) the subtraction of the giant spectrum.
 The first number in each of the right panels shows the date of
 observations in YYMMDD format, and the second the orbital phase.
 Note that these phases were calculated with the orbital period of
 of Fekel et al. (\cite{fek}), but the zero phase refers to the
 inferior conjunction of the RG.}
 \label{spec}
 \end{figure*}

For more than 30 years the HC in \object{T~CrB} has been
considered as a main sequence star (Kraft \cite{kraft}; Kenyon \&
Garcia \cite{kg}). Investigations over the last 10 years,
however, suggest that the accreting object must be a white
dwarf (WD). First, \cite{sel}  analyzed long-term ultraviolet (UV)
spectral observations of \object{T~CrB} and concluded that the HC
is a WD. Belczynski \& Mikolajewska (\cite{bm}) analyzed the
ellipsoidal variability of the RG and suggested the presence of a WD.
In a study of recurrent novae with giant secondaries, 
Anupama \& Mikolajewska (\cite{am}) also found that the HCs in these 
systems are WDs embedded in the dense wind of the RG.
Finally, Hachisu \& Kato (\cite{kato}) successfully modeled
recurrent novae outbursts as thermonuclear runaway on a massive
WD. However, no system parameters determination based on the
radial velocities (RVs) of the two components has been
published since Kraft (\cite{kraft}). Recently, Hric et al.
(\cite{hric}) re-analyzed the RVs of Kraft (\cite{kraft}) and
concluded that the hot component in \object{T~CrB} is a WD. However their
analysis is still based on Kraft's seven measurements of the HC RV.

In this paper we present an analysis of spectral H$\alpha$ and 
photoelectric $UBV$ 
observations of \object{T~CrB} obtained during  the last decade.
This is the first investigation of this object where the RG
contribution is subtracted and the "cleaned" H$\alpha$ profile is
measured. Based on our H$\alpha$ radial velocity measurements we
obtain a new orbital solution. We also discuss the long-term
variability of \object{T~CrB}.

\section{Observations and spectra processing}

 \begin{table}[t]
 \caption[]{The equivalent width EW and flux of the H$\alpha$ line, 
 its equivalent width after the giant spectrum
 subtraction EW(-gM), the factors $f$, and the radial velocity data of the
 HC, RG and central dip. The accuracy of the measurement is
 $\sim10$\% in EW (mainly due to the uncertainty of the continuum
 placement), $\sim20-40$\% in the flux (continuum placement
 and uncertainty of the continuum flux around H$\alpha$) and
 EW(-gM) (uncertainty of $f$) and $\sim10-15$\,km\,s$^{-1}$ in the
 radial velocities.}

\includegraphics*[width=8.8cm]{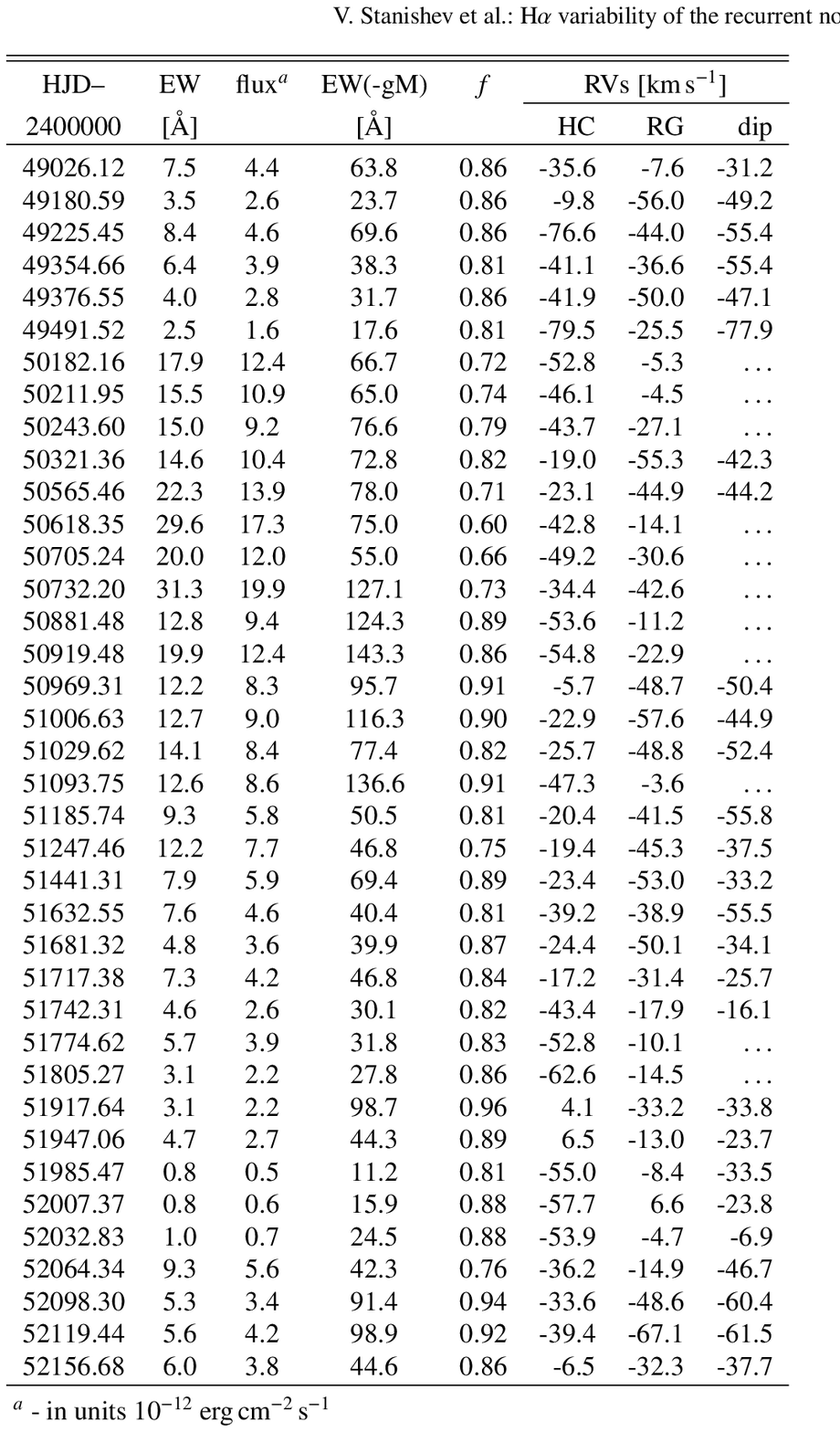}
\label{obs}
\end{table}

 The star \object{T~CrB} was observed during 64 nights between 
 February 1993 and September 2001 with the Coud\'e spectrograph of
 the 2.0~m RCC telescope of the Bulgarian National Astronomical 
 Observatory "Rozhen". The spectra cover  
 $\sim100$\,\AA\ or $\sim200$\,\AA\ around H$\alpha$, with 
 resolution of $\sim0.2$\,\AA\,pixel$^{-1}$.
 In some nights we have also obtained spectra covering the 
\ion{He}{ii}\,$\lambda$4686 emission line.
 Usually 2 or 3 exposures, of 15-20 minutes each, were taken to improve the
 sensitivity and to avoid problems with cosmic ray hits.
 The spectra were reduced in the standard way including bias removal, 
 flat-field
 correction, wavelength calibration and correction for the Earth's motion.
 The spectra obtained within each observational season
 (with a typical duration of 2-4 days)  were averaged to improve
 the signal-to-noise ratio. 38 mean H$\alpha$ spectra were obtained.
 They are plotted in the left panels of Fig.\,\ref{spec}.
We have also obtained new photoelectric $UBV$ observations 
with the 60-cm telescope at Rozhen Observatory 
(Zamanov et al. 2004, in preparation).

 The main source of the
continuum around H$\alpha$ is the RG and the contribution of the
HC is rather small, of the order of 5-15\%. Because of this and since 
the H$\alpha$ emission line is usually weak, the profile is strongly affected by the
absorption lines of the RG. Therefore, to measure the RVs of the
H$\alpha$ emission line accurately one should first subtract the spectrum of
the giant. To do this we used the following procedure. The M4\,III
star \object{HD134807} (Buscombe \cite{mk1}) was observed with the
same instrumental setup as \object{T~CrB} and served as a
template. The template was continuum-normalized and shifted to
zero RV. Once H$\alpha$ was masked, the RVs of the giant were
obtained by convolving the \object{T~CrB} spectra with the
template (Tonry \& Davis \cite{tonry}). The template was then
shifted to the RV of each spectrum, multiplied by a number $f$
between 0 and 1 and subtracted from the \object{T~CrB} spectra.
$f$ was chosen to minimize the scatter of the residuals.
The width of the absorption lines in the spectra of both
\object{T~CrB} and the template is dominated by the spectral
resolution rather than the rotational broadening. As a result 
the absorption lines have almost equal widths; however,
 in some cases the line widths are different. In those cases 
the average absorption line widths in the template and \object{T~CrB} spectra
were estimated by an autocorrelation function. 
Before the subtraction, the template or the spectrum was broadened
by convolving it with a Gaussian to equalize the line
widths. The "cleaned" H$\alpha$ emission line profiles and the factors $f$ 
used are shown in the right panels of Fig.\,\ref{spec} and in 
Table\,\ref{obs}, respectively.

 \begin{figure}[t]
  \centering
  \includegraphics*[width=8.8cm]{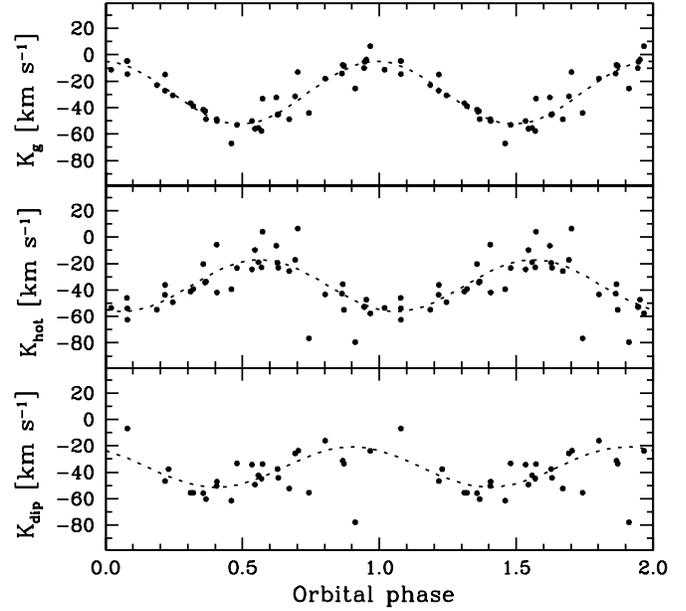}
  \caption{The RVs of the giant, HC and the dip folded with
  the ephemeris of Fekel et al. (\cite{fek}).
  The best fits are shown with dotted lines.}
  \label{rvs}
 \end{figure}

\section{Radial velocities}
\label{rv}

The radial velocities of the H$\alpha$ emission line were measured by fitting 
the line wings with a Gaussian function. 
We fitted only the outer parts of the line which 
 are not contaminated by the central asymmetry/dip.
Figure\,\ref{spec} shows that even after subtraction of the giant's
spectrum the H$\alpha$ line is double-peaked.  When possible,
the RV of the dip was also measured. A simple fitting with two
Gaussian functions was used -- one broad with a positive amplitude
for the emission line and another narrow with a negative amplitude
for the dip. Figure\,\ref{rvs} shows the RVs of the giant,
H$\alpha$ emission line and the dip folded with the ephemeris of
Fekel et al. (\cite{fek}) $T_0=JD 2447918.62+227\fd5687\,E$. The
RVs were fitted with the function $V_{\rm
r}=\gamma+K\cos[2\pi(\phi-\phi_0)]$ and the obtained parameters
are given in Table\,\ref{rvsfit}.

The RV of the giant follows the ephemeris of Fekel et al. (\cite{fek})
very well. The RV of the H$\alpha$ wings changes in anti-phase with that
of the giant. This suggests that the location of the emission is on the line
connecting the two stars, on the side of the mass center which is opposite 
to the giant.  Most likely, the region where H$\alpha$ emission line is
formed is centered close to the HC.
The observed ellipsoidal variability
(Shahbaz et al. \cite{sha}; Yudin \& Munari \cite{yum}; 
Belczynski \& Mikolajewska \cite{bm}) 
suggests that the giant fills its
Roche lobe, which provides a possibility for mass flow through the inner 
Lagrangian point and an AD formation. In this case the most plausible 
assumption is that the bulk of the H$\alpha$ flux is emitted by the outer part
of the AD. There is also other indirect evidence 
for the presence of an AD in this system (\cite{sel}; 
Anupama \& Mikolajewska \cite{am}; Hachusu
\& Kato \cite{kato}). Since the AD orbits around the
mass center of the system together with the HC, the RVs of the emission 
line wings are likely to trace the motion of the HC. 
We will then use the radial velocity curves to estimate 
the component masses.  Note, however, the small difference in
the $\gamma$ velocity of the HC and RG, and that $\phi_0$ of the HC
RVs is not exactly 0.5. This is likely to be a result of the
presence of additional emission components which most probably
arise in the ionized parts of the giant's wind. 
Such emission components can be easily seen in
Fig.\,\ref{spec} (the panels for 960608, 970618 and 980811).

\begin{table}[t]
  \centering
  \caption[]{The fit parameters to the H$\alpha$ RV data.}
 \begin{tabular}{lccc}
  \hline
  \hline
  \noalign{\smallskip}
          & red giant & hot component & central dip \\
  \hline
  \noalign{\smallskip}
$\gamma$ [km\,s$^{-1}$] & -28.7 (1.4) & -36.7 (1.6) & -36.1 (2.3) \\
$K$  [km\,s$^{-1}$]     & 23.7 (1.8) & 19.5 (2.1) & 15.2 (3.0) \\
$\phi_0$  & -0.009 (0.013) & 0.563 (0.019) & -0.10 (0.03) \\
$\sigma$ [km\,s$^{-1}$] & 8.2 & 9.0 & 10.0 \\
   \hline
   \end{tabular}
   \label{rvsfit}
\end{table}

If the double-peaked shape of H$\alpha$ is a result of emission
from an AD seen under moderate-to-high inclination, then the RV
of the central dip should be in phase with that of the line
wings. Figure\,\ref{rvs} shows that the RV of the dip is modulated
with the orbital period but follows more closely that of the
giant, which argues against the AD origin. The central dip in 
the emission profiles of the symbiotic stars is thought 
to originate mainly in the giant's wind. In the symbiotic star 
\object{EG And} the RV of the dip also changes in phase with that of the 
giant (Munari \cite{mun}). Thus, the central dip in the 
H$\alpha$ emission profile of \object{T~CrB} 
originates most probably in the giant's wind.

\section{Orbital solution}

 \begin{figure}[t]
   \centering
   \includegraphics*[width=8.8cm]{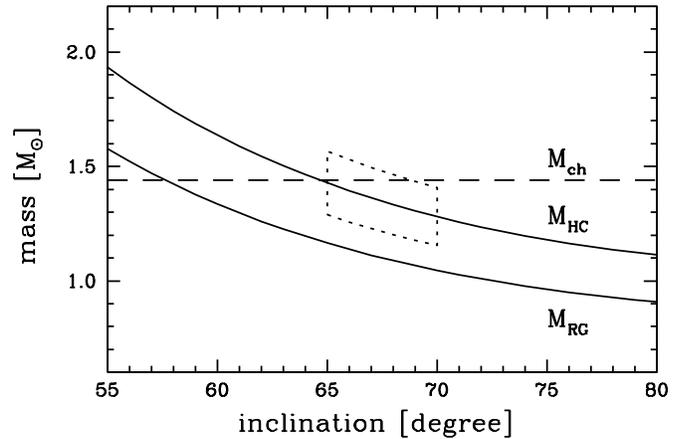}
   \caption{The masses of the components of \object{T~CrB} as a function of the inclination.
   The dotted curve shows the most likely position of the HC and the
   $i-mass$ space.}
   \label{mass}
 \end{figure}

Having the RV amplitudes of the two components we
can estimate their masses. The velocity amplitude of the giant in
\object{T~CrB} was accurately determined by Kenyon \& Garcia
(\cite{kg}) -- $K_{\rm RG}=23.32\,\pm0.16$ km\,s$^{-1}$ and was also
recently revised to $K_{\rm RG}=23.89\,\pm0.17$ km\,s$^{-1}$
by Fekel et al. (\cite{fek}). Our giant's velocity data are in full 
agreement with the results of these authors. But since our measurements are 
noisier, they lead to larger uncertainty of $K_{\rm RG}$. Thus, for 
determination of the system's parameters we prefer to use $K_{\rm RG}$ of Fekel et
al. (\cite{fek}). Assuming a zero eccentricity we obtain the following 
parameters: $M_{\rm
HC}\,\sin^3i=1.06\,\pm0.10M_\odot$ and $M_{\rm
RG}\,\sin^3i=0.87\,\pm0.18M_\odot$.

The masses of the two components as a function of the system inclination
are shown in Fig.\,\ref{mass}. Our
HC velocities allow solutions with $M_{\rm HC}$ below
the Chandrasekhar limit when the inclination is $i\geq65^\circ$. To
obtain a more precise estimation of the  masses a tighter
constraint on the inclination is needed. The HC contributes little in the 
visible wavelengths but it gives almost the whole UV flux. Therefore, 
if the HC was eclipsed this would be best 
observed in the UV. The UV data, however, does not show any hint
of eclipse (\cite{sel}) and it can provide an upper
limit.  We assume that the giant fills its Roche
lobe.  We  have then numerically calculated the projection of the RG 
in the orbital plane at zero phase for several values of $i$ between 
60$^\circ$ and 80$^\circ$, and component mass ratio 
$q=M_{\rm RG}/M_{\rm HC}=K_{\rm HC}/K_{\rm RG}=0.82\,\pm0.10$.
We found that the center of the HC is not eclipsed when
$i\leq70^\circ$. The upper limit depends also on the
size of the UV emitting region. Its size, however,  should be  very
small compared to the radius of the HC's Roche lobe  because the most likely
UV emitting region is the WD and the inner part of the AD (\cite{sel}).
Thus, we conclude that the upper limit of the inclination is
$\sim70^\circ$. The dotted lines in Fig.\,\ref{mass} mark the
limits of the inclination and $\pm1\sigma$ of the HC mass.
Recently, Hachisu \& Kato (\cite{kato}) have modeled the outburst light
curve of \object{T~CrB} and obtained the WD mass $M_{\rm
WD}\simeq1.37\,\pm0.01M_\odot$. In terms of our orbital solution
this corresponds to $i\simeq67^\circ$ and for this inclination we
have $M_{\rm WD}\simeq1.37\,\pm0.13M_\odot$ and $M_{\rm
RG}\simeq1.12\,\pm0.23M_\odot$.

 \begin{figure*}[!ht]
    \centering
   \includegraphics*[width=12cm]{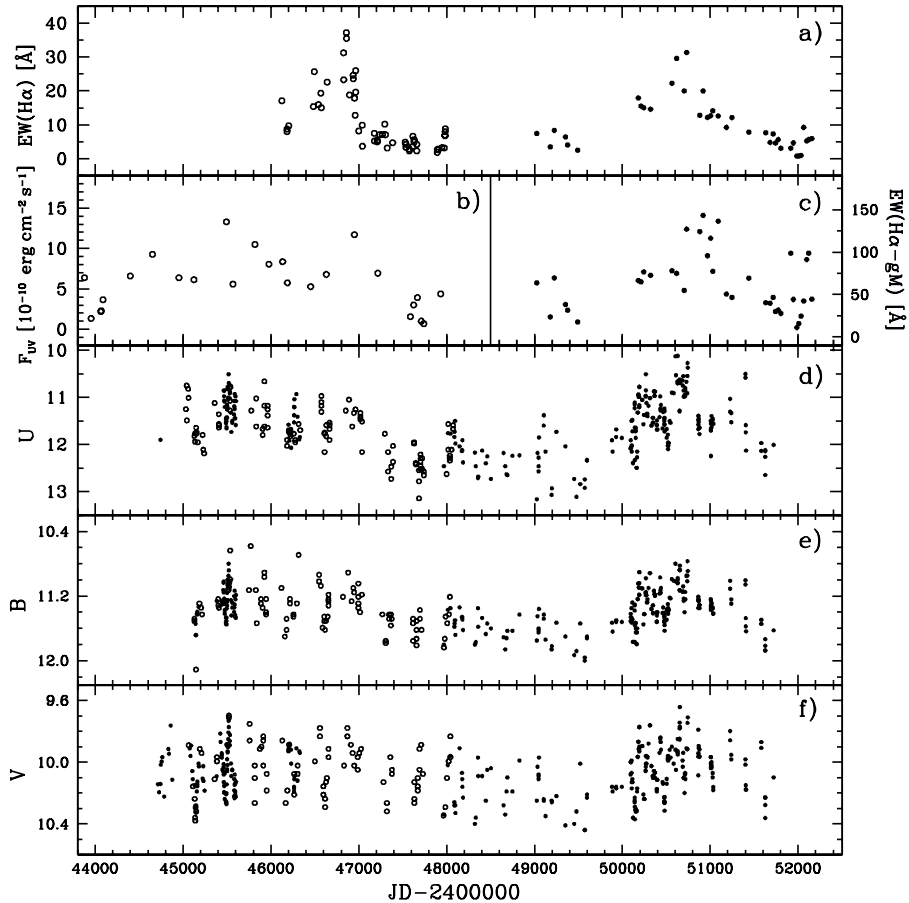}
   \caption{\ {\bf a)} H$\alpha$ equivalent width;
   {\bf b)} total UV flux between 1250 and 3200\,\AA;
   {\bf c)} H$\alpha$ equivalent width
    after the subtraction  of the RG spectrum; {\bf d)}
   $U$ magnitudes, {\bf e)} $B$ magnitudes and 
   {\bf f)} $V$ magnitudes. See text for 
   the sources of the different quantities.}
   \label{long}
 \end{figure*}

\section{Long-term variability}

In Fig.\,\ref{long} we show our observations of \object{T~CrB}
along with those published. Figure\,\ref{long}$a$ shows our
EW(H$\alpha$) measurements (dots) and those of Anupama \& Prabhu (\cite{anu})
(open circles). Figures\,\ref{long}$b$ and $c$ show the total UV flux
between 1250 and 3200\,\AA\ ($F_{\rm UV}$; taken from \cite{sel}) and 
EW(H$\alpha$) measured 
from our spectra after the subtraction of the RG contribution 
(hereafter EW(H$\alpha$-gM)), respectively. In the photometry panels 
(Figs.\,\ref{long}$d-f$) the dots show the standard $UBV$ magnitudes
(Raikova \& Antov \cite{ra}; Lines et al. \cite{lines}; Hric et al.
\cite{hric}; Sherrington \& Jameson \cite{sj}; Zamanov \& Zamanova
\cite{zz}; our new data) and the open circles the instrumental ones 
(Oskanian \cite{osk}; Luthardt \cite{luth}). The data of
Luthardt were presented only in graphical
form as a difference between \object{T~CrB} and the comparison star
which was not specified. 
We estimated the magnitudes of \object{T~CrB} from the figure of
Luthardt (\cite{luth}) in the following way.
On the computer screen we pointed with the cursor on each data point 
and read its coordinates. These "computer screen" coordinates were 
converted to Julian Dates and magnitudes by applying the 
necessary transformations. Then, the data
were shifted to the scale of the standard magnitudes by using the
intervals when they overlap with the $UBV$ observations.
We have also obtained an estimation of the 
H$\alpha$ flux by using the
measured EW(H$\alpha$) and the published $R$-band magnitudes. We
assume that the $R$-band flux represents the continuum flux at 
the position of H$\alpha$ and that the variations in $R$ 
are entirely due
to the ellipsoidal variability. We fitted the $R$ data of Mikolajewski et
al. (\cite{mik}) and Hric et al. (\cite{hric}) with a function
consisting of three sines with periods of $P_{\rm orb}$, $P_{\rm
orb}/2$ and $P_{\rm orb}/3$. With the coefficients obtained we have
calculated the $R$ magnitudes at the moments of the spectral
observations. Then, Bessell's (\cite{bess}) calibration of the
$R$-band was used to calculate the H$\alpha$ flux. Because the 
H$\alpha$ flux variations repeat very closely those of EW(H$\alpha$),
the H$\alpha$ flux is given only in Table\,\ref{obs} 
(only for our spectra). We note that the assumption that the
$R$-band flux represents the flux in the continuum around
H$\alpha$ introduces the largest uncertainty in the calculated
H$\alpha$ flux. The reason is that the $R$ filter covers several 
strong molecular bands, and is
placed where the flux in the RG spectrum increases strongly toward
the long wavelengths.

As seen from Fig.\,\ref{long}  our observations cover an interval 
when both the photometric data and EW(H$\alpha$) vary greatly.
The long-term variability has amplitude  $\sim2.0$, 1.0 and 
0.3 mag in $U$, $B$ and $V$ bands, respectively. 
EW(H$\alpha$) increased from $\sim5$\,\AA\ to  $\sim30$\,\AA.
EW(H$\alpha$-gM) also increased, but its maximum 
is delayed with respect to the maximum of EW(H$\alpha$) and
appears at JD2450950. It is seen from Figs.\,\ref{long}$d$ and $e$ 
that the maximum of EW(H$\alpha$-gM) coincides with 
the rapid decrease of the $U$ and $B$-band fluxes. This might 
explain the observed peculiarity. It is also 
seen from Fig.\,\ref{long} that most of the UV observations have been 
obtained when \object{T~CrB} was in a high state. It is quite evident,
however, that during the observations around JD2444000
and JD2447700 \object{T~CrB} was in a low state.
The average increase of $F_{\rm UV}$ is $\sim3.3$ and 
the corresponding increase of the $U$-band flux is  
$\sim2.6$ ($\sim1$ mag).

\begin{figure}
   \centering
   \includegraphics*[width=8.8cm]{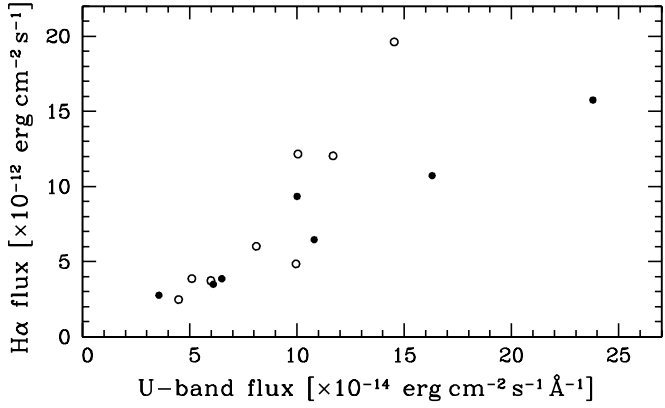}
   \caption{The yearly averaged H$\alpha$ flux {\it vs.}
   the $U$-band flux. Dots represent out data and open circles
   those of  Anupama \& Prabhu (\cite{anu}).}
   \label{flux}
 \end{figure}

Anupama \& Mikolajewska (\cite{am}) noticed that the increase
of EW(H$\alpha$) in the late 1980s correlated with the increase of 
UV and $U$ fluxes. Our new observations also show  similar behavior.
 To verify the strong visual impression
for correlation of the data shown in Fig.\,\ref{long} we
performed a correlation analysis. The small 
number of UV observations required interpolation when calculating
the correlations involving $F_{\rm UV}$. For the correlation 
between H$\alpha$ and $U$-band fluxes we used yearly averaged values.
(EW(H$\alpha$-gM) was not used here).
The computed Pearson's correlation coefficient and Spearman's rank 
correlation coefficient are in all cases larger than 0.7 with the
significance of the correlation always higher than 99\%.
The correlation analysis confirmed that all
the data in Fig.\,\ref{long} are indeed strongly correlated. 
As an example we show H$\alpha$ {\it vs.}
 $U$-band flux in Fig.\,\ref{flux}. Our data and those of Anupama \& Prabhu (\cite{anu})
follow different tracks, but the correlation is apparent.

\begin{figure*}
    \centering
    \includegraphics*[width=12cm]{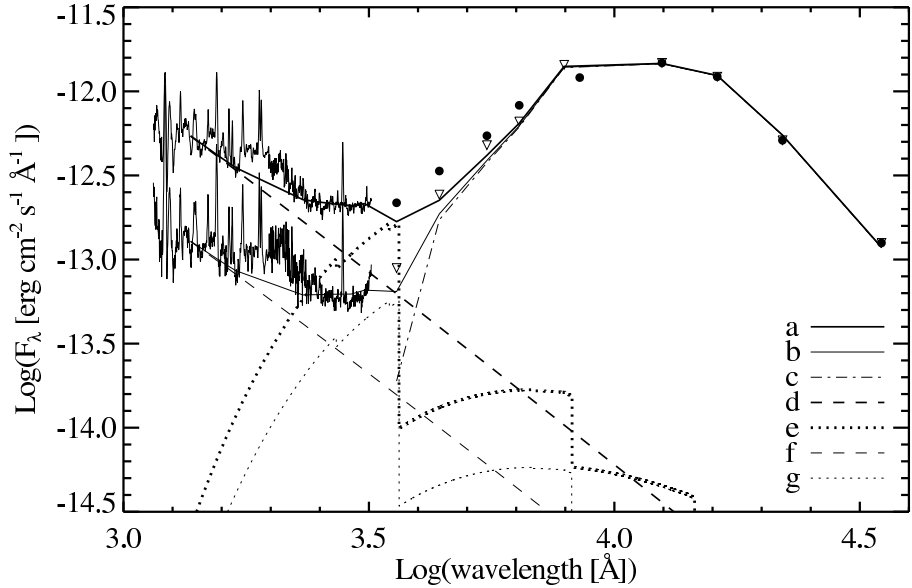}
   \caption{UV-optical-IR energy distribution of \object{T~CrB}.
    The representative UV spectra are related to its low and high states.
    The triangles and the filled dots indicate the photometric fluxes 
    at low and high states, respectively. 
    Lines {\bf a} and {\bf b} show the model spectrum of the system 
    in high and low state, respectively. Line {\bf c} connects the 
    synthetic fluxes of a red giant. Lines {\bf d} and {\bf e} indicate 
    the continia of the accretion disc and the nebula in high state. 
    Lines {\bf f} and {\bf g} show the same components for the low state.}
    \label{en}
  \end{figure*}

The correlated variations in the continuum can be also seen
in Fig.\,\ref{en} where we present the UV-optical-IR spectral 
energy distribution (SED) of \object{T~CrB} in low and high states. 
The points are a compilation (average) from the photometric data of 
Shahbaz et al. (\cite{sha}), Kamath \& Ashok (\cite{ka}),
Yudin \& Munari (\cite{yum}), Munari et al. (\cite{mun1}),
Hric et al. (\cite{hric}), Mikolajewski et al. (\cite{mik}), 
Zamanov \& Zamanova (\cite{zz}) and our measurements. 
A representative high-state UV spectrum of \object{T~CrB} was
obtained by averaging the low-resolution spectra taken by  
the International Ultraviolet  Explorer ($IUE$) between
JD2444300 and JD2447300 when the system was in high state.
The rest of the $IUE$ spectra were used to obtain 
the average low-state UV spectrum.
All fluxes
were corrected for interstellar reddening of $E_{B-V}=0.15$ (Cassatella et al. 
\cite{cas}) using O'Donnell's (\cite{red}) interstellar extinction law 
and $R_V=3.1$.

The data in Figs.\,\ref{long}$b$ and \ref{en} show that the bolometric 
luminosity of the hot component varies greatly on time scale 
of about 200 days, which is not a typical feature of objects
whose activity is powered by thermonuclear runaways.
Moreover, \cite{sel} demonstrated that the bolometric luminosity 
of the hot  component is low ($\sim100L_\odot$) compared to some other symbiotic 
systems (e.g., \object{AG Peg}, \object{AG Dra}, \object{Z And} ) 
whose activity is supposed to be due to hydrogen burning and that even in the high 
state most of the hot component's light is emitted in the UV.
All these characteristics  suggest that the system is accretion powered and
the hot component is most
probably
the innermost part of an optically thick accretion disc and/or  
the boundary layer between the disc and the WD's surface as  
was already proposed by \cite{sel}. This idea supports our previous 
assumption based on the behavior of the H$\alpha$ radial velocity 
(Sect. 3) that the bulk of the  energy of this line is probably
emitted by the outer part of an AD.

According to the AD theory 
(e.g., Warner \cite{cvs}),  there exists a critical value
$\dot{M}_{\rm cr}$ of the mass accretion rate $\dot{M}$, so that
if $\dot{M}>\dot{M}_{\rm cr}$ the disc is in a steady state, being hot 
and optically thick with effective temperature $T_{\rm
eff,hot}\propto \dot{M}^{1/4}M_{\rm WD}^{1/4}r^{-3/4}$ 
($T_{\rm eff,hot}>10^4$\,K). Here $r$ is the distance to the WD.
If $\dot{M}<\dot{M}_{\rm cr}$ the disc is
cool and optically thin with $T_{\rm eff,cool}\simeq5-6\times10^3$\,K.
With the parameters of \object{T~CrB}
$\dot{M}\sim10^{-8}\,M_\odot$\,yr$^{-1}$ (\cite{sel}) and
$M_{\rm WD}\simeq1.37$, only the inner part of the disc with radius
$\sim 1 R_\odot$ is in the hot state, while the rest of the disc is
cool. In the case of accretion at a high rate onto a very massive WD, as
in \object{T~CrB}, the temperature of the innermost part of the 
disc should be $\sim1-2\times10^5$\,K. It will thus emit mostly 
in the UV and can be the HC. In this case variations of $\dot{M}$  
can easily account for its observed variability.

The symbiotic systems containing a hot compact object change their 
energy distribution appreciably during active phase because of the variation 
of the temperature of the compact object. 
Figure\,\ref{en} shows that the UV energy distribution 
of \object{T CrB} tends not to vary
during its transition from low to high states. 
\cite{sel}  notice 
a correlation of the slope of the continuum with the total UV flux. 
However, using the values given in their
Table 1, we found that the correlation, if any, is very weak.
Thus, 
the  modeling of the continuum requires nearly the same temperature
of the hot component in low and high states.  We tried 
to model the continuum of \object{T CrB} assuming the presence
of an optically thick AD. The 
IR region was fitted with synthetic fluxes of a model spectrum of 
a red giant with  $T_{\rm eff}=3500$\,K, $M=1M_\odot$ (P. Hauschildt, private 
communication) and $R=75R_\odot$, filling 
its Roche lobe. The UV and visual regions were fitted with 
an emission of an AD and gas environment. 
 We  adopted an electron temperature of
10000\,K according to \cite{sel}. Thus,
the only free parameters are the mass accretion rate and the relative
scaling between the three components.

To determine the nebular continuum,  
the ionization state of helium has to be known. The dominant state of ionization 
can be estimated from the ratio of the emission measures of the neutral and ionized 
helium based on the fluxes of some lines of \ion{He}{i} and the \ion{He}{ii}\,$\lambda$4686 
line. Since the electron temperature of 10000 K is comparatively low we assumed  
that \ion{He}{i} lines are determined only by recombination. 
Our observations of the \ion{He}{ii}\,$\lambda$4686 line show that
most of the time this line was absent
and only occasionally appeared very weak. The 
line was also not presented in the spectrum in July 1986, as can 
be seen in Fig.\,2 of \cite{sel}. 
According to Iijima (1990) in the period 1981-1987 \object{T CrB} has 
had prominent emission lines of \ion{He}{i} whereas its \ion{He}{ii}\,$\lambda$4686  
line was only temporarily  observed. In the spectra taken in Apr--May 1997
during the recent high state, 
Anupama \& Mikolajewska (\cite{am}) detected the \ion{He}{ii}\,$\lambda$4686  
line, but it was by a factor of 2 weaker than the \ion{He}{i} lines.
At an electron temperature of 10000\,K and electron density of 
$10^6$~cm$^{-3}$ if the number of He$^+$ and He$^{++}$ ions is the same, 
the intensity of the \ion{He}{ii}\,$\lambda$4686 line
exceeds the intensities of the \ion{He}{i} lines by one order of magnitude
(e.g. Pottasch \cite{pot}).
This was probably not the case during the period of Iijima's observations.
Thus, we accepted the
dominant helium ionization state \ion{He}{ii} in all occasions and assumed that the nebular 
continuum is mostly continuum emission 
of hydrogen and neutral helium. 
We adopted a helium abundance of 0.1 (Nussbaumer \& Vogel \cite{nv}) and 
distance to the system of 1.3 kpc (Patterson \cite{pat}).

The result of the modeling is shown in Fig.\,\ref{en}. 
To determine the level of the UV continuum of \object{T CrB} is a 
complicated task -- 
the spectrum is dominated by humps caused by blended emission lines. We
fitted the bottoms of the troughs between those humps. The
largest discrepancies between the fit and the spectrum are in the short-wavelength region
 and around 2200\,\AA\ 
where the sensitivity of $IUE$ is low. 
It is seen, however, that the shape of the fit to the UV spectrum
in the two states is nearly the same. The observed optical fluxes in 
both states are systematically higher than the fit. This indicates the 
presence of component(s) which are not taken onto account in our model.
This additional light might be due to density and 
temperature inhomogeneities in the outer disc or/and in the other nebular 
formations in the system.

 The modeling gave  an accretion rate of 
$4\times10^{-9}\,M_\odot$\,yr$^{-1}$  for the low state and 
$3\times10^{-8}\,M_\odot$\,yr$^{-1}$  for the high state in agreement 
with \cite{sel}. Note also that at these accretion 
rates the disc temperature exceeds 10$^5$\,K. At such temperatures
the scope of the UV continuum is nearly constant.

We obtained appreciable emission measures of $2.7\times10^{59}$\,cm$^{-3}$ and 
$7.4\times10^{59}$\,cm$^{-3}$  for the low and high states, determined 
from strong nebular UV and optical continuum. On the other hand the 
Balmer emission lines are weak (H$\alpha$ exceeds the level of the 
continuum by a factor of not greater than 5-6). The symbiotic
systems with an extended surrounding nebula (\object{AG Dra}, \object{Z And}) 
have prominent Balmer emission lines exceeding the continuum by a factor 
of up to 40-50, together with their intensive nebular 
continuous spectrum. In the case of \object{T CrB} the weakness of the 
Balmer emission lines 
can be understood if they are assumed to be formed in a gas medium with a 
great optical depth, which absorbs the dominant part of the photons. The 
optical depth is great when the column density of the emitting gas is 
high.  Anupama \& Mikolajewska (\cite{am}) came to the
conclusion that the hot component of the symbiotic
recurrent novae is most probably a white dwarf $+$
accretion disc embedded in an optically thick envelope
formed by the wind of the M giant secondary. This model provides high 
column density and contains an AD as well. So, the weak Balmer lines of 
\object{T CrB} can be qualitatively explained by the assumption of
an origin in a binary system containing an AD. The investigation of 
the RV of the H$\alpha$ line showed that the bulk of its energy is 
emitted by an area  around the hot component -- probably the outer part
of an AD. This model is supported also by our suggestion for the explanation of
the weak Balmer emission lines of \object{T CrB}. The modeling 
of \object{T CrB}'s SED suggests that the 
UV spectrum appears flat mostly because of the strong nebular contribution.
However, it is also possible that this is partly a result of seeing 
the AD at a relatively high inclination as suggested by \cite{sel}.

The evidence for an AD and high
$\dot{M}$ provides good conditions for formation of intensive 
high-speed wind emerging from the innermost hot parts of the disc 
(see Mauche \& Raymond \cite{wind} for a discussion of disc wind 
in the case of 
cataclysmic variables). Consequently the nebular spectrum of the 
system may appear also in the winds of the disc and
the cool giant in addition of the disc itself. 
Shore \& Aufdenberg (\cite{sho}) showed that the giant's wind is variable.
Further evidence for this can be found in the variations of the 
central dip of the H$\alpha$ emission line -- most of the time the 
dip is well pronounced, but sometimes it is highly variable and is 
almost absent (Fig.\,\ref{spec}). 

If an AD is present in \object{T CrB} system, its 
activity may be a result of variation of the 
accretion rate $\dot{M}$.
There are two ways to increase $\dot{M}$: (1) to increase the mass transfer 
rate $\dot{M}_{\rm tr}$  from the giant and (2) to increase temporarily the 
mass flow through the AD. The former may be a result of 
the magnetic activity of the
giant as suggested by Soker (\cite{sok}). The later mechanism corresponds to 
the outbursts observed in the dwarf novae and is related to 
the so-called disc instability. In the context
 of symbiotic stars the disc instability was studied by Duschl 
(\cite{dus}), but assuming that the accreting object is an
$\sim1M_{\odot}$ main-sequence star rather than a WD.
Briefly, in disc the instability model 
$\dot{M}_{\rm tr}$ is constant and less than $\dot{M}_{\rm cr}$, i.e. 
the disc is in the cool state most of the time. Outburst is observed 
when $\dot{M}$ (locally in the disc) increases above $\dot{M}_{\rm cr}$ and 
the AD rapidly approaches the hot state
(for more details see Warner \cite{cvs}). 
In both cases the disc luminosity increases because of the dependence 
of the temperature on the mass accretion rate: 
$T_{\rm eff,hot}\propto\dot{M}^{1/4}$.

\section{Summary and conclusions}

The radial velocities of the H$\alpha$ emission line of the recurrent nova
\object{T~CrB} were precisely measured after subtraction of the red giant 
spectrum, thus obtaining the unbiased H$\alpha$ profile. 
Based on our 38 new radial velocity  measurements we obtain
a new solution for the system parameters. The
solution indicates that at a system inclination
$i>65^\circ$ the mass of the hot component is below the Chandrasekhar
limit and thus the hot component in \object{T~CrB} is most likely a massive
white dwarf. The most likely values of the system inclination and the
component masses are $i\simeq67^\circ$,  $M_{\rm
WD}\simeq1.37\,\pm0.13M_\odot$ and $M_{\rm
RG}\simeq1.12\,\pm0.23M_\odot$.

The radial velocity of the central dip of the H$\alpha$ emission 
line profile changes
almost in phase with that of the red giant. This suggests that the
central dip originates from the giant's wind.

During our observations \object{T~CrB} showed a period of 
activity when the system brightness increased by
$\sim2.0$, 1.0 and 0.3 mag in $U$, $B$ and $V$
bands, respectively. The H$\alpha$ equivalent width also increased
by a factor of $\sim6$ reaching $\sim30$\,\AA\ and the line 
had a complex multi-component structure.
The phasing of the modulation of H$\alpha$ emission line radial velocity 
shows that the 
emitting region is gravitationally connected to the hot component
of the system and most probably related to the outer part of an 
accretion disc. We argue that a number of other characteristics of  \object{T~CrB}
are also in support of this supposition.
The long-term variability is then assumed to be caused by variations 
of the
mass accretion rate, possibly due to changes of the mass transfer 
rate from the companion or disc instability.
This is also supported by our success in modeling the 
UV-optical-IR SED of \object{T CrB} with a three-component model: an 
optically thick accretion disc, a nebular continuum and a red giant. However, to construct a
consistent picture and to understand the underlying physical
processes in \object{T~CrB} better, simultaneous UV and
optical spectral observations during future active phases are
needed. Space observatories offering simultaneous optical/UV
spectroscopic capabilities like Kronos (Peterson et al.
\cite{pet}) and WSO/UV (Wamsteker \& Ponz \cite{wam}) would be a
good solution.


\end{document}